\documentclass{article}

\usepackage[sans]{dsfont}
\usepackage[mathscr]{euscript}
\usepackage{cite}

\usepackage{amssymb}          
\usepackage{amsmath}          

\newtheorem{prop}{Proposition}
\newtheorem{defin}{Definition}
\newtheorem{thm}{Theorem}
\newtheorem{cor}{Corollary}
\newtheorem{lemma}{Lemma}

\newcommand{\ket}[1]{|#1\rangle}
\newcommand{\bra}[1]{\langle #1|}

\newcommand{\Hi}{\mathcal{H}}

\newcommand{\supp}{\textrm{supp}}

\newcommand{\beq}{\begin{equation}}
\newcommand{\eeq}{\end{equation}}
\newcommand{\beqa}{\begin{eqnarray}}
\newcommand{\eeqa}{\end{eqnarray}}
\newcommand{\beqan}{\begin{eqnarray*}}
\newcommand{\eeqan}{\end{eqnarray*}}

\newcommand{\qed}{\hfill $\Box$ \vskip 2ex}
\newcommand{\proof}{\noindent {\em Proof.}}

\newcommand{\Li}{\mathcal{L}}

\renewcommand{\ker}{\rm{ker}}
\renewcommand{\span}{\rm{span}}

\newcommand{\tr}{\textrm{trace}}

\begin{document}
\title{Stabilizing Entangled States with\\ Quasi-Local Quantum
Dynamical Semigroups}
\author{Francesco Ticozzi\thanks{Dipartimento di Ingegneria dell'Informazione, 
Universit\`a di Padova,
via Gradenigo 6/B, 35131 Padova, Italy ({\tt ticozzi@dei.unipd.it}) and
Department of Physics and Astronomy,
Dartmouth College, 6127 Wilder Laboratory, Hanover, NH 03755, USA} 
{ }and{ }%
  Lorenza Viola\thanks{Department of Physics and Astronomy,
Dartmouth College, 6127 Wilder Laboratory, Hanover, NH 03755, USA,
({\tt lorenza.viola@dartmouth.edu})}}
\maketitle
\abstract{We provide a solution to the problem of determining whether
a target pure state can be asymptotically prepared using dissipative
Markovian dynamics under fixed locality constraints. Beside recovering
existing results for a large class of physically relevant entangled states, 
our approach has the advantage of providing an explicit stabilization test 
solely based on the input state and constraints of the problem.  Connections 
with the formalism of frustration-free parent Hamiltonians are discussed, as
well as control implementations in terms of a switching
output-feedback law.}
%

\section{Introduction}

While uncontrolled couplings between a quantum system of interest and its 
surrounding environment are responsible for unwanted non-unitary evolution 
and decoherence, it has also been long acknowledged that suitably engineering and 
exploiting the action of the environment  may prove beneficial in a number 
of applications across quantum control and quantum information processing
(Poyatos {\em et al.} 1996, Beige {\em et al}. 2000, Lloyd \& Viola 2001). 
It is well known, in particular, that open-system dynamics are instrumental in 
control tasks such as robust quantum state preparation and rapid purification, 
and both open-loop and quantum feedback methods have been extensively 
investigated in this context
(Combes {\em et al.} 2008, Wiseman \& Milburn 2009,  Ticozzi \& Viola 2009, 
Schirmer \& Wang 2010, Combes {\em et al.} 2010), including recent extensions to 
engineered quantum memories (Pastawski {\em et al.} 2010) and pointer states in the 
non-Markovian regime (Khodjasteh {\em et al.}  2011).

Remarkably, it has also been recently shown that it is in principle possible to 
design dissipative Markovian dynamics so that non-trivial strongly correlated 
quantum phases of matter are prepared in the steady state (Diehl {\em et al.} 
2008) or the output of a desired quantum algorithm is retrieved as the asymptotic 
equilibrium (Verstraete {\em et al.} 2009).  From a practical standpoint, scalability 
of such protocols for multipartite systems of increasing size is a key issue, as 
experimental constraints on the available control operations may in fact limit the 
set of attainable states. Promising results have been obtained by Kraus 
{\em et al.} (2008) for a large class of entangled pure states, showing that 
Markovian dissipation acting non-trivially only on a {\em finite} 
maximum number of subsystems is, under generic conditions, sufficient to 
generate the desired state as the unique ground state of the resulting evolution.  
As proof-of-principle methodologies for engineering dissipation are becoming an 
experimental reality (Barreiro {\em et al.} 2011, Krauter {\em et al.} 2011), it is 
important to obtain a more complete theoretical characterization of the set of 
attainable states under constrained control resources, as well as to explore 
schemes for synthesizing the required dissipative evolution.

Building on our previous analysis (Ticozzi \& Viola 2008, 2009, Ticozzi {\em et al.} 
2010),  in this work we address the problem of determining whether a target
pure state of a finite-dimensional quantum system can be prepared employing 
``quasi-local'' dissipative resources with respect to a {\em fixed} locality notion 
(see also Yamamoto's contribution to this volume for related results on 
infinite-dimensional Markovian Gaussian dissipation). We provide a 
stabilizability analysis under locality-constrained 
Markovian control, including a direct test to verify whether a desired entangled 
pure state can be asymptotically prepared.  In addition to recovering existing 
results within a system-theoretic framework, our approach has the
important advantage of using only two inputs: the desired state
(control task) and a specified locality notion (control constraints),
without requiring a representations of the state in the
stabilizer, graph, or matrix-product formalisms.

\section{Problem definition and preliminary results}

\subsection{Multipartite systems and locality of QDS generators}

We focus on quantum evolutions driven by a (time-independent)
Markovian Master Equation (MME) (Gorini {\em et al.} 1976, Lindblad 1976, 
Alicki \& Lendi 1987) in Lindblad form ($\hbar \equiv 1$): \beq
\label{eq:lindblad}\dot\rho(t)={\cal
L}(\rho(t))=-i[H,\,\rho(t)]+\sum_k \Big(
L_k\rho(t)L^\dag_k-\frac{1}{2}\{L_k^\dag L_k,\,\rho(t)\} \Big), \eeq
specified in terms of the Hamiltonian $H=H^\dagger$ and a finite set
of noise (or Lindblad) operators $\{L_k\}$. We are interested on the
asymptotic behavior of MMEs in which the operators $H,\{L_k\}$ satisfy
locality constraints.  More precisely, let us consider a multipartite
system ${\cal Q}$, composed of $n$ (distinguishable) subsystems,
labeled with index $a=1,\ldots,n$, with associated $d_a$-dimensional
Hilbert spaces $\Hi_a$.  Thus, $\Hi_{\cal Q}=\bigotimes_{a=1}^n\Hi_a.$
Let ${\mathfrak B}(\Hi)$ and ${\mathfrak D}(\Hi)$ denote the sets of
linear operators and density operators on $\Hi$, respectively.  It is
easy to show (see {\em e.g.} Ticozzi \& Viola 2008, proof of Theorem 2)
that the semigroup generated by Eq. \eqref{eq:lindblad} is factorized
with respect to the multipartite structure, that is, the dynamical
propagator
\[{\cal T}_t := e^{{\cal L}t}=\bigotimes_{a=1}^n {\cal
T}_{a,t}\quad\forall t\geq 0,\] with ${\cal T}_{a,t}$ a CPTP map on
${\mathfrak B}(\Hi_{a}),$ if and only if the following conditions
hold:

\vspace*{1mm}

(i) Each $L_k$ acts as the identity on all subsystems except (at most)
one;

\vspace*{1mm}

(ii) $H=\sum_aH_a,$ where each $H_a$ acts as the identity on all
subsystems except (at most) one.

\vspace*{1mm}

\noindent 
This motivates the following definitions: we say that a {\em noise
operator $L_k$ is local} if it acts as the identity on all subsystems
except (at most) one, and that a {\em Hamiltonian $H$ is local} if it
can be written as a sum of terms with the same property. However, it is 
easy to verify that if a
semigroup associated to local operators admits a unique stationary
pure state, the latter must necessarily be a product state. Thus, in
order for the MME \eqref{eq:lindblad} to admit stationary entangled
states, it is necessary to weaken the locality constraints.

We shall allow the semigroup dynamics to act in a non-local way only
on certain subsets of subsystems, which we call {\em
neighborhoods}. These can be generally specified as subsets of the set
of indexes labeling the subsystems:
\[{\cal N}_j\subseteq\{1,\ldots,n\}, \quad j=1,\ldots, M.\]
In analogy with the strictly local case, we say that a {\em noise
operator $L$ is Quasi-Local (QL)} if there exists a neighborhood ${\cal
N}_j$ such that:
\[L= L_{{\cal N}_j}\otimes I_{\bar{\cal N}_j},\] 
where $L_{{\cal N}_j}$ accounts for the action of $L$ on the
subsystems included in ${\cal N}_j$, and $I_{\bar{\cal
N}_j}:=\bigotimes_{a\notin{\cal N}_j}I_a$ is the identity on the
remaining subsystems.  Similarly, a {\em Hamiltonian is QL} if it
admits a decomposition into a sum of QL terms: \[H=\sum_jH_j, \quad
H_j=H_{{\cal N}_j}\otimes I_{\bar{\cal N}_j}.\] 
\noindent 
A MME will be called QL if {\em both} its Hamiltonian and noise operators
are QL. It is well known that the decomposition into Hamiltonian and
dissipative part of \eqref{eq:lindblad} is not unique: nevertheless,
the QL property remains well defined since the freedom in the
representation does not affect the tensor structure of $H$ and
$\{L_k\}$.  The above way of introducing locality constraints is very
general and encompasses a number of specific notions that have been
used in the physical literature, notably in situations where the
neighborhoods are associated with sets of nearest neighbors sites on a
graph or lattice, and/or one is forced to consider Hamiltonian and
noise generators with a weight no larger than $t$ (so-called $t$-body
interactions), see also Kraus {\em et al.} 2008, Verstraete {\em et al.}
2009. 

We are interested in states that can be prepared (or, more precisely,
stabilized) by means of MME dynamics with QL operators.  Recall that
an invariant state $\rho$ for a system driven by \eqref{eq:lindblad}
is said to be {\em Globally Asymptotically Stable (GAS)} if for every
initial condition $\rho_0$ we have
\[\lim_{t\rightarrow \infty} e^{{\cal L}t}(\rho_0)= \rho. \] 
In particular, following Kraus {\em et al.} (2008), the aim of this paper
is to characterize pure states that can be rendered GAS by {\em purely
dissipative dynamics}, for which the state is ``dark''. More
precisely:

\begin{defin} A pure state 
$\rho_d=\ket{\Psi}\bra{\Psi}\in{\mathfrak D}(\Hi_{\cal Q}),$ is
Dissipatively Quasi-Locally Stabilizable (DQLS) if there exist QL
operators $\{D_k\}_{k=1,\ldots,K}$ on $\Hi_{\cal Q},$ with
$D_k\ket{\Psi}=0,$ for all $k$ and $D_k$ acting non-trivially on (at
most) one neighborhood, such that $\rho_d$ is GAS for
\beq\label{eq:lindblad0} \dot\rho={\cal L}_D[\rho]=\sum_k \Big(
D_k\rho D_k^\dag-\frac{1}{2}\{D_k^\dag D_k,\,\rho\} \Big).\eeq
\end{defin}
We will provide a test for determining whether a state is DQLS, and in
doing so, we will also show how assuming a {\em single} QL noise operator
for each neighborhood does not restrict the class of stabilizable
states. From now on, we thus let $K\equiv M,$ and $D_k\equiv 
D_{{\cal N}_k}\otimes I_{\bar{\cal N}_k}$.

We begin by noting that if a pure state is factorized, then we can
realize its tensor components ``locally'' with respect to its
subsystems (see Ticozzi \& Viola 2008, Ticozzi {\em et al.} 2010) for 
stabilization of arbitrary quantum states in a given system with simple 
generators, involving a single noise term).  Thus, we can iteratively reduce 
the problem to subproblems on disjoint subsets of subsystems, until the
states to be stabilized are either entangled, or completely
factorized.  A preliminary result is that the DQLS property is
preserved by arbitrary Local Unitary (LU) transformations, of the form
$U=\bigotimes_{a=1}^{n} U_a$. In order to show this, the following
lemma is needed.

\begin{lemma}\label{unitaryrot} Let  
${\cal L}$ denote the Lindblad generator associated to operators
$H,\{L_k\}.$ Then for every unitary operator $U$ we have 
\beq 
\label{L1} 
U{\cal L} [U^\dag\rho U]U^\dag={\cal L}'[\rho]\eeq 
\noindent 
where ${\cal L}'$ is the semigroup generator associated to $H'=U H
U^\dag,$ $L_k'=U L_k U^\dag,$ and \beq
\label{L2} 
Ue^{{\cal L}t}[U^\dag\rho_0U]U^\dag= e^{{\cal L}'t}[\rho_0], \quad
\forall t \ge0. \eeq
\end{lemma}

\noindent 
Identity \eqref{L1} is easily proven by direct computation, while
\eqref{L2} follows directly from the properties of the (matrix) exponential.
The desired invariance of the QL stabilizable set under LU
transformation follows:

\begin{prop} If $\rho$ is DQLS and U is LU, then 
$\rho'=U\rho U^\dag$ is also DQLS.
\end{prop}
\proof Assume that the generator ${\cal L}$ associated to QL operators
$\{D_k\}$ stabilizes $\rho.$ Since $\rho$ is GAS, for any initial
condition $\rho_0$ we may write 
\beq
\label{first} 
\rho'=\lim_{t\rightarrow +\infty} Ue^{{\cal L} t}[\rho_0]U^\dag=
\lim_{t\rightarrow +\infty} U(e^{{\cal L}
t}[U^\dag\rho_0U])U^\dag.\eeq 
\noindent 
By applying Lemma \ref{unitaryrot}, it suffices to show that each
$D'_k=U D_k U^\dag$ is QL.  Since $D_k = D_{{\cal N}_{k}}\otimes
I_{\bar{\cal N}_{k}},$ we have $D'_k=U (D_{{\cal N}_{k}}\otimes
I_{\bar{\cal N}_{k}}) U^\dagger = \left(U_{{\cal N}_{k}} D_{{\cal N}_{k}}
U^\dag_{{\cal N}_{k}}\right)\otimes I_{\bar{\cal N}_{k}},$ where $U_{{\cal
N}_{k}}:=\bigotimes_{\ell\in{\cal N}_{k}} U_\ell.$ Hence $D'_k$ is QL.
\qed

\subsection{QDS for unconstrained stabilization}

We next collect some stabilization results that do not directly
incorporate any locality constraint, but will prove instrumental to
our aim.  Let $\Hi_S:=\span\{{\ket{\Psi}}\}.$ Given Corollary 1 in
Ticozzi \& Viola 2008, $\rho_d$ is invariant if and only if
\beq\label{eq:nonrobust} L_{k}=\left[
\begin{array}{c|c}
 L_{S,k} & L_{P,k} \nonumber \\\hline 0 & L_{R,k}
\end{array}
\right], \quad iH_P-\frac{1}{2}\sum_kL_{S,k}^\dag L_{P,k}=0, \eeq
where we have used the natural block representation induced by the partition
$\Hi_{\cal Q}=\Hi_S\oplus\Hi_S^\perp$ and labeled the blocks as
\[ X=\left[
\begin{array}{c|c}
  X_{S} & X_P     \\\hline
  X_Q  &  X_R   
\end{array}
\right]. \] 
\noindent 
Assume $\rho_d$ to be invariant. Hence $\ket{\Psi}$ must be a common
eigenvector of each $L_k$.  Call the corresponding eigenvalue
$\ell_k \equiv L_{S,k}.$ By Lemma 2 in Ticozzi \& Viola (2008), the MME generator
is invariant upon substituting $L_k$ with $\tilde L_k=L_k-\ell_k I,$
and $\tilde H=H + i\sum_k (\ell_k^*L_k-\ell_kL_k^\dag).$ In this way,
we have $\tilde L_{S,k}=0$ for all $k,$ so that $\tilde H_P$ must be
zero in order to fulfill the above condition.  Thus, $\tilde H$ is
block-diagonal, with $\ket{\Psi}$ being an eigenvector with eigenvalue
$h\equiv \tilde H_S.$ Using this representation for the generator, we can
let $L_{S,k}=\ell_k=0,$ and $H_P=0=H_Q.$ This further
motivates the use of noise operators $D_k$ such that $D_k\ket{\Psi}=0$
in the DQLS definition.

\begin{lemma}
\label{cor2}  
An invariant $\rho_d=\ket{\Psi}\bra{\Psi}$ is GAS for the MME 
\eqref{eq:lindblad} if there are no invariant common (proper)
subspaces for $\{L_k\}$ other that $\Hi_S=\span\{\ket{\Psi}\}.$
\end{lemma} \proof By Lemma 8 and Theorem 9 in
Ticozzi \& Viola 2009, $\rho_d$ is GAS if and only if there are no
other invariant subspaces for the dynamics. Given the conditions on
$L_k$ for the invariance of a subspace, $\rho_d$ is GAS as long as 
there are no other invariant common subspaces for the matrices $L_k.$ 
\qed

Based on the above characterization, in order to ensure the DQLS 
property it suffices to find operators $\{D_k\}_{k=1}^K\subset{\mathfrak
B}(\Hi_{\cal Q})$ such that $\Hi_S$ is the {\em unique} common (proper)
invariant subspace for the $\{D_k\}.$

\section{Characterization of DQLS states}

\subsection{Main result}

Our main tool for investigation will be provided by the {\em reduced
states} that the target state $\rho_d$ induces with respect to the given local
structure.  Let us define: 
\beq\label{redstate} \rho_{{\cal
N}_k}=\tr_{\bar{\cal N}_k}(\rho_d),\eeq where $\tr_{\bar{\cal N}_k}$
indicates the partial trace over the tensor {\em complement} of the
neighborhood ${\cal N}_k,$ namely $\Hi_{\bar{\cal
N}_k}=\bigotimes_{a\notin{\cal N}_k}\Hi_a.$ The following Lemma
follows from the properties of the partial trace:

\begin{lemma}\label{rhovsrhok} 
$\supp(\rho_d)\subseteq \bigcap_k\supp(\rho_{{\cal N}_k}\otimes
I_{\bar{\cal N}_k}).$
\end{lemma}
\proof From the spectral decomposition $\rho_{{\cal
N}_k}=\sum_qp_q\Pi_q$, we can construct a resolution of the identity
$\{\Pi_q\}$ such that
$\rho_{{\cal N}_k}\otimes 
I_{\bar{\cal N}_k}=\sum_{q}p_{q}\Pi_q\otimes I_{\bar{\cal N}_k},$
where, by definition of the partial trace,
$p_{q}=\tr(\rho_d\Pi_q\otimes I_{\bar{\cal N}_k}).$ If $p_{\hat q}=0$
for some $\hat q$, then it must be $\rho_d(\Pi_{\hat q}\otimes
I_{\bar{\cal N}_k})=0$ and therefore $\supp(\rho_d)\perp
\supp(\Pi_{\hat q}\otimes I_{\bar{\cal N}_k}).$ Thus, 
$\supp(\rho_d)\subseteq\bigcup_q\supp(p_q\Pi_q\otimes I_{\bar{\cal
N}_k})=\supp(\rho_{{\cal N}_k}),$ for all $k.$ \qed
\noindent 
Let us now focus on QL noise operators $D_k= D_{{\cal
N}_k}\otimes I_{\bar{\cal N}_k}$ such that $D_k\ket{\Psi}=0.$  

\begin{lemma}
\label{supp} Assume that a set $\{D_k\}$ makes 
$\rho_d=\ket{\Psi}\bra{\Psi}$ DQLS. Then, for each $k,$ we have
$\supp(\rho_{{\cal N}_k})\subseteq \ker(\tilde D_{{\cal N}_k}).$
\end{lemma}

\proof Since $\ket{\Psi}$ is by hypothesis in the kernel of each
$D_k$, with respect to the decomposition
$\Hi_{Q}=\Hi_S\oplus\Hi_S^\perp$ every $D_k$ must be of block form:
\[D_k=\left[\begin{array}{cc} 0 & D_{P,k} \\ 0 & D_{R,k}\end{array}\right],\]
which immediately implies
$D_k\rho_d D_k^\dag =0 .$
\noindent 
It then follows that $\tr_{\bar{\cal N}_k}(D_k\rho_d D_k^\dag)=0=
\tr_{\bar{\cal N}_k}( D_{{\cal N}_k}\otimes I_{\bar{\cal
N}_k}\rho_d  D_{{\cal N}_k}^\dag\otimes I_{\bar{\cal N}_k})$.
Therefore, it also follows that 
$D_{{\cal N}_k}\rho_{{\cal N}_k} D_{{\cal N}_k}^\dag=0.  $
If we consider the spectral decompositon $\rho_{{\cal
N}_k} \equiv \sum_jq_j\ket{\phi_j}\bra{\phi_j}$, with $q_j>0,$
the latter condition implies that, for each $j,$ $\tilde D_{{\cal
N}_k}\ket{\phi_j}\bra{\phi_j}\tilde D_{{\cal N}_k}^\dag=0$.  Thus, it
must be $\supp(\rho_{{\cal N}_k})\subseteq \ker(\tilde D_{{\cal
N}_k})$, as stated. \qed

\begin{thm}
\label{mainthm} 
A pure state $\rho_d=|\psi\rangle\langle \psi|$ is DQLS if and only if
\beq\label{GAScond}\supp(\rho_d)=\bigcap_k\supp(\rho_{{\cal
N}_k}\otimes I_{\bar{\cal N}_k}) \equiv \bigcap_k {\mathcal H}_{{\cal
N}_k}.\eeq
\end{thm}

\proof Given Lemmas 3 and 4, for any set $\{D_k\}$ that make $\rho_d$ DQLS
we have:
\[\supp(\rho_d)\subseteq \bigcap_k\supp(\rho_{{\cal N}_k}
\otimes I_{\bar{\cal N}_k})\subseteq\bigcap_k\ker(D_{{\cal
N}_k}\otimes I_{\bar{\cal N}_k}).\] 
\noindent 
By negation, assume that $\supp(\rho_d)\subsetneq
\bigcap_k\supp(\rho_{{\cal N}_k}\otimes I_{\bar{\cal N}_k})$. Then
there would be (at least) another invariant state in the intersection
of the kernels of the noise operators, contradicting the fact that
$\rho_d$ is DQLS. Thus, a necessary condition for $\rho_d$ to be GAS
is that $\supp(\rho_d)=\bigcap_k\supp(\rho_{{\cal N}_k}\otimes
I_{\bar{\cal N}_k}).$ Conversely, if the latter condition is
satisfied, then for each $k$ we can construct operators $\hat D_{{\cal
N}_k}$ that render each $\supp(\rho_{{\cal N}_k})$ GAS on $\Hi_{{\cal
N}_k}$ (see {\em e.g.} Ticozzi \& Viola 2008, 2009, Ticozzi {\em et al.} 2010 
for explicit constructions). Then $\bigcap_k\ker(\hat
D_{{\cal N}_k}\otimes I_{\bar{\cal N}_k})=\supp(\rho_{{\cal N}_k}),$
and there cannot be any other invariant subspace. By Lemma \ref{cor2},
$\rho_d$ is hence rendered GAS by QL noise operators.  \qed

\subsection{An equivalent characterization: QL parent Hamiltonians}
 
Consider a QL Hamiltonian $H=\sum_k H_k,\quad H_k=H_{{\cal
N}_k}\otimes I_{\bar{\cal N}_k}.$ A pure state $\rho_d =|\Psi
\rangle\langle \Psi |$ is called a {\em frustration-free ground state}
if
\[\langle \Psi |H_k | \Psi \rangle=\textrm{min}\; 
\lambda(H_k),\quad\forall k,\] 
\noindent 
where $\lambda(\cdot)$ denotes the spectrum of a matrix. A QL
Hamiltonian is called a {\em parent Hamiltonian} if it admits a {\em
unique} frustration-free ground state (Perez-Garcia {\em et al.} 2007). 

Suppose that a pure state admits a QL parent Hamiltonian $H$. Then 
the QL structure of the latter can be naturally used to derive a
stabilizing semigroup: it suffices to implement QL operators $L_k$
that stabilize the eigenspace associated to the minimum eigenvalue of
each $H_k$. In view of Theorem 1, it is easy to show that this
condition is also necessary:

\begin{cor}\label{parent}
A state $|\Psi\rangle$ is DQLS if and only if it is the ground state
of a QL parent Hamiltonian.
\end{cor}

\proof Without loss of generality we can consider QL Hamiltonians
$H=\sum_k H_k,$ where each $H_k$ is a projection. Let $\rho_d$ be
DQLS, and define $H_k:=\Pi^\perp_{{\cal N}_k}\otimes I_{\bar{\cal
N}_k}$, with $\Pi^\perp_{{\cal N}_k}$ being the orthogonal projector
onto the orthogonal of the support of $\rho_{{\cal N}_k}$, that is,
$\Hi_{{\cal N}_k}\ominus \supp({\rho_{{\cal N}_k}})$. Given Theorem
\ref{mainthm}, $|\Psi\rangle $ is the unique pure state in
$\bigcap_k\supp(\rho_{{\cal N}_k}\otimes I_{\bar{\cal N}_k}),$ and
thus the unique state in the kernel of all the $H_k$. Conversely, if a
QL parent Hamiltonian exists, to each $H_k$ we can associate an $L_k$
that asymptotically stabilizes its kernel. A single operator per
neighborhood is in principle always sufficient (see Ticozzi \& Viola 
2008, 2009, Ticozzi {\em et al.} 2010 for explicit
constructions and examples of $L_k$ stabilizing a desired subspace).
\qed

The above result directly relates our approach to the one pursued in
Kraus {\em et al.} 2008, Verstraete {\em et al.} 2009, and a few remarks 
are in order.  In these works it has been shown that Matrix Product States
(MPS) are QL stabilizable, up to a condition (so-called {\em injectivity})
that is believed to be generic (Perez-Garcia {\em et al.} 2007). MPS states 
that allow for a compact representation (that is, in the corresponding 
``valence-bond picture'', those with a small bond dimension) are of
key interest in condensed matter as well as quantum information
processing (Verstraete {\em al.} 2006, Perez-Garcia {\em et al.} 2007, 
Perez-Garcia {\em et al.} 2008). However, {\em
any} pure state admits a (canonical) MPS representation if 
sufficiently large bond dimensions are allowed, suggesting that
arbitrary pure states would be DQLS. The problem with this reasoning
is that the locality notion that is needed in order to allow
stabilization of a certain MPS is in general {\em induced by the state
itself}. The number of elements to be included in each neighborhood is
finite but need not be small: while this is both adequate and
sufficient for addressing many relevant questions in many-body physics
(where typically a thermodynamically large number of subsystems is
considered), engineering the dissipative process may entail
interactions that are not easily available in experimental
settings. For this reason, our approach may be more suitable for
control-oriented applications.  It is also worth noting that the
injectivity property is sufficient but not necessary for the state to admit a 
QL parent Hamiltonian (an example on a two-dimensional 
lattice is provided in Perez-Garcia {\em et al.} 2007). 
Once the locality notion is fixed, our test for DQLS can be performed 
{\em irrespective} of the details of the MPS representation, and it is thus  
not affected by whether the latter is injective or not (rather, our DQLS test may be
used to output a QL parent Hamiltonian if so desired).

\subsection{Examples}

\noindent{\bf $\bullet$ GHZ-states and W-states.--} Consider an 
$n$-qubit system and a target GHZ state $\rho_{\text{GHZ}} =|\Psi
\rangle\langle \Psi |$, with $\ket{\Psi}\equiv
\ket{\Psi_{\text{GHZ}}}=(\ket{000\ldots 0}+\ket{111\ldots
1})/\sqrt{2}.$ Any reduced state on any (nontrivial) neighborhood is
an equiprobable mixture of $\ket{000\ldots 0}$ and
$\ket{111\ldots 1}.$ It is then immediate to see that
\[\textrm{span}\{\ket{000\ldots 0}, 
\ket{111\ldots 1}\}\subseteq\bigcap_k\supp(\rho_{{\cal N}_k}\otimes
I_{\bar{\cal N}_k}),\] and hence $\rho_{\text{GHZ}}$ is {\em not}
DQLS.  In a similar way, for any $n$ the W state $\rho_{\text{W}}
=|\Psi \rangle\langle \Psi|$, with $\ket{\Psi}\equiv
\ket{\Psi_{\text{W}}}=(\ket{100\ldots 0}+\ket{010\ldots 0}+\ldots
+\ket{000\ldots 1})/\sqrt{n}$ has reduced states that are statistical
mixtures of $\ket{000\ldots 0}$ and a smaller W state
$\ket{\Psi_{\text{W}'}}$, of the dimension of the neighborhood. Thus,
\[\textrm{span}\{\ket{000\ldots 0}, \ket{\Psi_{\text{W}'}} \}
\subseteq\bigcap_k\supp(\rho_{{\cal N}_k}\otimes I_{\bar{\cal
N}_k}),\] and $\rho_{\text W}$ is {\em not} DQLS (except in trivial
limits, see also below). Note that for arbitrary $n$, both
$\rho_{\text{GHZ}}$ and $\rho_{\text{W}}$ are known to be (non-injective) 
MPS with (optimal) bond dimension equal to two.

\vspace{2mm}

\noindent{\bf $\bullet$ Stabilizer and graph states.--} A large class
of states does admit a QL description, and in turn they are
DQLS. Among these are stabilizer states, and general graph
states. Here the relevant neighborhoods are those that include all the
nodes connected to a given one by an edge of the graph. The details
are worked out in Kraus {\em et al.} 2008. Notice that GHZ states {\em
are} indeed graph states, but only associated to star (or completely
connected) graphs. Hence, relative to the locality notion naturally
induced by the graph, any central node has a neighborhood which
encompasses the whole graph, rendering the constraints trivial.

\vspace{2mm}

\noindent{\bf $\bullet$ DQLS states beyond graph states.--} Consider a
4-qubit system arranged on a linear graph, with (up to) 3-body
interactions. The two neighborhoods ${\cal N}_1=\{1,2,3\}, {\cal
N}_2=\{2,3,4\}$ are sufficient to cover all the subsystems, and
contain all the smaller ones. The state $\rho_{T} =|\Psi
\rangle\langle \Psi |$ with
\[\ket{\Psi}\equiv \ket{\Psi_T}=(\ket{1100}+\ket{1010}+\ket{1001}+\ket{0110}+
\ket{0101}+\ket{0011})/\sqrt{6},\]
\noindent 
is {\em not} a graph state, since if we measure any qubit in the
standard basis, we are left with W states on the remaining subsystems,
which are known not to be graph states. In contrast, Proposition 9 of
Hein {\em et al.} 2004 ensures that the conditional reduced states
for a graph state would have to be graph states as well. Nonetheless,
by constructing the reduced states and intersecting their supports one
can establish directly that $\ket{\Psi}_T$ is indeed DQLS.

\section{Switched feedback implementation}

From Theorem \ref{mainthm} it follows that a DQLS state can be
asymptotically prepared provided we can engineer QL noise operators
$D_k= D_{{\cal N}_k}\otimes I_{\bar{\cal N}_k}$ that stabilize
the support of each reduced state $\rho_{{\cal N}_k}$ on each
neighborhood.  Restricting to $\Hi_{{\cal N}_k},$ we must have 
$D_{{\cal N}_k}=\left[\begin{array}{cc} 0 &  D_{P,k} \\ 0 &
D_{R,k}\end{array}\right],$ with the blocks 
$D_{P,k}, D_{R,k}$ such that the support of $\rho_{{\cal N}_k}$
is attractive, that is, such that no invariant subspace is contained
in its complement.  Following the ideas of Ticozzi \& Viola (2009), 
Ticozzi {\em et al.} (2010), a natural explicit choice is to
consider noise operators with the following structure:
\begin{equation}
\label{uppdiag}
     D_{P,k}=
     \begin{bmatrix}
      0 & 0 & \cdots & 0 \\
      \vdots & 0 & \cdots & 0 \\
      \ell_1 & 0 & \cdots & 0 \\
     \end{bmatrix},\;
     D_{R,k}=
     \begin{bmatrix}
      0 & \ell_2 & 0 & 0 \\
      0 & 0 & \ell_3 & \ddots \\
      \vdots &  & \ddots & \ddots \\
\end{bmatrix}.
\end{equation}

If the above QL Lindblad operators are not directly available for
open-loop implementation, a well studied strategy for synthesizing
attractive Markovian dynamics is provided by continuous measurements
and output feedback.  In the absence of additional dissipative
channels, and assuming perfect detection, the relevant \emph{Feedback
Master Equation} takes the form (Wiseman \& Milburn 2009):
\[ 
\dot{\rho}(t)= -i \left[H +H_c+ \tfrac{1}{2}(FM+M^\dag
F),\,\rho(t)\right] + L_f\rho (t)L_f^\dag-\frac{1}{2}\left \{L_f^\dag
L_f,\rho (t)\right\} , \] 
\noindent 
where $H_c$ is a time-independent control Hamiltonian, $F=F^\dagger$
and $M$ denote respectively the feedback Hamiltonian and the measurement 
operator, and $L_f := M -i F$.  Necessary and sufficient conditions for the
existence of open- and closed- loop Hamiltonian control that
stabilizes a desired subspace have been provided in Ticozzi \& Viola 2008, 
2009.

In order to exploit the existing techniques in the current
multipartite setting, it would be necessary to implement measurements
and feedback in each neighborhood. If the measurement operators do not
commute, however, one would have to carefully scrutinize the validity
of the model and the consequences of ``conflicting'' stochastic
back-actions when acting {\em simultaneously} on overlapping
neighborhoods.  These difficulties can be bypassed by resorting to a
{\em cyclic switching of the control laws}.  Consider a DQLS state
$\rho_d$ and the family of generators $\{{\cal L}_k\}_{k=1}^{M},$
${\cal L}_k [\rho]=D_k\rho D_k^\dag-\frac{1}{2}\{D_k^\dag D_k,\rho\}$,
with $D_k$ such that $\textrm{supp}(\rho_{{\cal N}_k}\otimes
I_{\bar{\cal N}_k})$ is the unique invariant subspace for ${\cal
L}_k$. Define a switching interval $\tau\geq 0$ and the cyclic
switching law $j(t)= \lfloor {t}/{\tau M} \rfloor + 1$.  We can then
establish the following:

\begin{thm}
\label{switching} 
There exists QL $\{D_k\}$ such that $\rho_d$ is GAS for the switched
evolution ${\cal L}_{j(t)}.$ \end{thm}

\proof Consider the trace-preserving, completely-positive maps ${\cal
T}_j(\rho)=e^{{\cal L}_j\tau}[\rho]$. It is easy to see that
$\rho_d$ is invariant for each ${\cal T}_j$: as a corollary of Theorem
1 in Bolognani \& Ticozzi 2010, it follows that $\rho_d$ is GAS if it is
the only invariant state for ${\cal T}={\cal T}_M\circ \cdots \circ
{\cal T}_1$.
Assume that $\rho$ is invariant for ${\cal T}$: then either it is fixed
for all ${\cal T}_k,$ which means that necessarily $\rho=\rho_d,$ or
there exists a periodic cycle. Since each ${\cal T}_j$ is a
trace-distance contraction (Alicki \& Lendi 1987), this means that each
map preserves the trace distance, that is, $\|{\cal
T}_j(\rho_d - \rho)\|_1=\|\rho_d - \rho\|_1.$ This would in turn imply
that each $ {\cal T}_j $ admits eigenvalues on the unit circle, and
hence each ${\cal L}_k$ would have imaginary ones. However, if we
choose $D_k$ as in Eq. \eqref{uppdiag}, in vectorized form the
Liouvillian generator reads
\[\hat\Li_k= D_k^{\dag T}\otimes D_k -\frac{1}{2} 
I\otimes D_k^\dag D_k-\frac{1}{2}(D_k^\dag D_k)^T\otimes I,\] 
\noindent 
which is an upper triangular matrix with eigenvalues either equal to
zero or $\{-(\ell_j^2+\ell_i^2)/2\}$. Therefore, for this choice
$\rho_d$ is the only invariant pure state state for ${\cal T}$ and
hence it is GAS.\qed

\section{Concluding remarks}

We have presented a characterization of DQLS pure states for {\em
fixed} locality constraints, from a control perspective. As a
byproduct of our main result, an easily automated algorithm for
checking DQLS states is readily devised. The necessary steps entail:
(1) calculating the reduced states on all the neighborhoods specifying
the QL notion; (2) computing their tensor products with the
identity on the remaining subsystems, and the relative supports; (3)
finding the intersection of these subspaces. If such intersection
coincides with the support of the target state alone, the latter is
DQLS. If so, we have additionally showed that the required Markovian
dynamics can in principle be implemented by {\em switching}
output-feedback control. While we considered
homodyne-type continuous-time feedback MME, the study of discrete-time 
strategies is also possible along similar
lines, see also  Bolognani \& Ticozzi (2010), Barreiro {\em et al.} (2011).

Our present results have been derived under two main assumptions: the
absence of underlying free dynamics, and the use of purely dissipative
control (no Hamiltonian control involved).  In case a drift internal
dynamics is present, the same approach can be adapted to determine
what can be attained by dissipative control.  When we additionally
allow for Hamiltonian control, one may employ the algorithm described
in Section III.B of  Ticozzi {\em et al.} (2011) to search for a viable QL
Hamiltonian when dissipation alone fails.  Nonetheless, in the
presence of locality constraints a more efficient design strategy may
be available: an in-depth analysis of these issues will be presented
elsewhere.

It is also worth noting that in various experimental situations the
available dissipative state preparation procedures involve two steps:
first, enact {\em local} noise operators that prepare a known pure
state that is factorized; next, use open-loop coherent control to
steer the system on the desired entangled target. The approach we
discussed here is believed to have an advantage in terms of the
overall robustness against initialization errors and finite-time
perturbations of the dynamics (Verstrate {\em et al.} 2009, Krauter {\em et al.} 
2011). While establishing rigorous robustness results requires further study, 
the actual answer is expected to depend on the physical implementation
and its characteristic time scales.  Lastly, the estimation of the
speed of convergence still present numerous challenges, most
importantly its optimization and a characterization of its scaling
with the number of subsystems involved.


\section*{Acknowledgements}
F.T. acknowledges support by the QUINTET and QFuture projects of the
University of Padova.


\begin{thebibliography}{1}

\bibitem[Alicki \& Lendi (1987)]{alicki-lendi}
Alicki, R. \& Lendi, K. 1987 \emph{Quantum Dynamical Semigroups and 
Applications}. Springer-Verlag, Berlin.

\bibitem[Barreiro {\em et al}. (2011)]{barreiro} Barreiro, J. T., Muller, M., Schindler, P., Nigg, D., 
Monz, T., Chwalla, M., Hennrich, M., Roos, C. F., Zoller, P. \& Blatt, R. 2011
An open-system quantum simulator with trapped ions. \emph{Nature}
{\bf 470}, 486--491.

\bibitem[Beige {\em et al}. (2000)]{beige} Beige, A., Braun, D., Tregenna, B. \&  Knight, P. L. 2000 
Quantum computing using dissipation to remain in a decoherence-free 
subspace. {\em Phys. Rev. Lett.} {\bf 85}, 1762--1765.

\bibitem[Bolognani \& Ticozzi (2010)]{bolognani-arxiv} Bolognani, S. \& Ticozzi, F. 2010  Engineering
stable discrete-time quantum dynamics via a canonical {QR}
decomposition.  {\em IEEE Trans. Aut. Contr.}  {\bf 55}, 2721--2734.

\bibitem[Combes {\em et al}. (2008)]{combes1} Combes, J., Wiseman, H. M. \& Jacobs, K. 2008 
Rapid measurement of quantum systems using feedback control.
{\em Phys. Rev. Lett.}  {\bf 100}, 160503:1--4. 

\bibitem[Combes {\em et al}. (2010)]{combes2} Combes, J., Wiseman, H. M. \& Scott, A. J. 2010 
Replacing quantum feedback with open-loop control and quantum 
filtering {\em Phys. Rev. A} {\bf 81}, 020301:1--4. 

\bibitem [Diehl {\em et al}. (2008)]{diehl} Diehl, S., Micheli, A., Kantian, A., Kraus, B., B\"{u}chler, H. P. \& 
Zoller, P. 2008 Quantum states and phases in driven open quantum systems 
with cold atoms. {\em Nature Phys.} {\bf 4}, 878--883. 

\bibitem[Gorini {\em et al}. (1976)]{gks} Gorini, V., Kossakowski, A. \& Sudarshan, E. G. C. 1976
Completely positive dynamical semigroups of $n$-level systems.  {\em
J. Math. Phys.}  {\bf 17}, 821--825.

\bibitem[Hein {\em et al}. (2004)]{hein-graphstates} Hein, M., Eisert, J. \& Briegel, H. J. 2004. 
Multiparty entanglement in graph states {\em Phys. Rev. A} {\bf  69}, 
062311:1--20.

\bibitem[Khidjasteh {\em et al}. (2011)]{Kaveh} Khodjasteh, K., Dobrovitski, V. V. \& Viola, L. 2011 
Pointer states via engineered dissipation. {\em Phys. Rev. A}  {\bf 84}, 
022336:1--21.

\bibitem[Kraus {\em et al}. (2008)]{kraus-entangled} Kraus, B., Diehl, S., Micheli, A., Kantian, A., 
B\"{u}chler, H. P. \& Zoller, P. 2008  Preparation of entangled states by
dissipative quantum Markov processes.  {\em Phys. Rev. A} 
 {\bf 78}, 042307:1--9.

\bibitem[Krauter {\em et al}. (2011)]{polzik} Krauter, H., Muschik, C. A., Wasilewski, W., 
Petersen, J. M., Cirac, J. I. \& Polzik, E. S 2011 Entanglement generated
by dissipation and steady state entanglement of two macroscopic
objects.  {\em Phys. Rev. Lett.}  {\bf 107}, 080503:1--4.

\bibitem[Lindblad (1976)]{lindblad} Lindblad, G 1976  On the generators of quantum
dynamical semigroups.  {\em Commun. Math. Phys.}  {\bf  48}, 119--130.

\bibitem[Lloyd \& Viola (2001)]{seth} Lloyd, S. \& Viola, L. 2001 Engineering quantum dynamics.
{\em Phys. Rev. A}  {\bf 65}, 010101:1--4.

\bibitem[Perez-Garcia {\em et al}. (2007)]{perez-MPS} Perez-Garcia, D., Verstraete, F., Wolf, M. M. \& 
Cirac, J. I. 2007 Matrix product state representations. {\em Quantum
Inf. Comput.}  {\bf 7}, 401--430.

\bibitem[Perez-Garcia {\em et al}. (2008)]{perez-PEPS} Perez-Garcia, D., Verstraete, F., Wolf, M. M. \& 
Cirac, J. I. 2008 PEPS as unique ground states of local Hamiltonians, {\em
Quant. Inf. Comp.}  {\bf 8}, 0650--0663.

\bibitem[Pastawski {\em et al}. (2011)]{pastawski} Pastawski, F., Clemente, L. \&  Cirac, J. I.  2011 
Quantum memories based on engineered dissipation.  
{\em Phys. Rev. A}  {\bf 83}, 012304:1--12.

\bibitem[Poyatos {\em et al}. (1996)]{Poyatos} Poyatos, J. F., Cirac, J. I. \& Zoller, P. 1996  
Quantum reservoir engineering with laser cooled trapped ions.
{\em Phys. Rev. Lett.}  {\bf 77}, 4728--4732. 

\bibitem[Schirmer \& Wang (2010)]{schirmer-markovian} Schirmer, S. G. \&  Wang, X. 2010
Stabilizing open quantum systems by markovian reservoir engineering.
{\em Phys. Rev. A}  {\bf 81}, 062306:1--14.

\bibitem[Ticozzi \& Viola (2008)]{ticozzi-QDS} Ticozzi F. \& Viola, L. 2008 Quantum Markovian
subsystems: Invariance, attractivity and control.  {\em IEEE
Trans. Autom. Contr.}  {\bf 53}, 2048--2063.

\bibitem[Ticozzi \& Viola (2009)]{ticozzi-markovian} Ticozzi F. \& Viola, L. 2009  Analysis and
synthesis of attractive quantum Markovian dynamics.  {\em Automatica}
{\bf 45}, 2002--2009.
 
\bibitem[Ticozzi {\em et al}. (2010)]{ticozzi-generic} Ticozzi, F., Schirmer, S. G. \& Wang, X. 2010
Stabilizing quantum states by constructive design of open quantum
dynamics.  {\em IEEE Trans. Autom. Contr.} {\bf 55}, 2901--2905.

\bibitem[Ticozzi {\em et al}. (2011)]{ticozzi-DID} Ticozzi, F., Lucchese, R., Cappellaro, P. \&
Viola, L. 2011 Hamiltonian control of quantum dynamical semigroups:
Stabilization and convergence speed. {\em IEEE Trans. Autom. Contr.}, 
in press. Preprint: arXiv:1101.2452v1.

\bibitem[Verstraete {\em et al}. (2006)]{perez-arealaw} Verstraete, F., Wolf, M. M., Perez-Garcia, D. \& 
Cirac, J. I. 2006 Criticality, the area law, and the computational power of
projected entangled pair states, {\em Phys. Rev. Lett.} {\bf 96}, 
220601:1--4.

\bibitem[Verstraete {\em et al}. (2009)]{verstraete-dissipation} Verstraete F., Wolf M. M. \& 
Cirac, J. I. 2009  Quantum computation and quantum-state engineering driven
by dissipation. {\em Nature Phys.} {\bf 5}, 633--636.

\bibitem[Wiseman \& Milburn (2009)]{wiseman-book} Wiseman H. M. \& Milburn G. J. 2009 \emph{Quantum
Measurement and Control}. Cambridge University Press, Cambridge.

\end{thebibliography}
\end{document}